\documentclass[useAMS,usenatbib]{mn2e}

\usepackage[latin1]{inputenc}
\usepackage{multirow}
\usepackage{graphicx}
\usepackage{multicol}
\usepackage{float}
\usepackage{color}
\usepackage{footnote}
\usepackage{amssymb}
\usepackage[figuresright]{rotating}
\usepackage{amsmath}
\usepackage{url}
\usepackage{rotating}
\usepackage{caption}
\newcommand{\apgt}{{\raise-.5ex\hbox{$\buildrel>\over\sim$}}}
\newcommand{\aplt}{{\raise-.5ex\hbox{$\buildrel<\over\sim$}}} 
\newcommand{\Mpc}{\rm\; Mpc}

\newcommand{\km}{\rm\; km}

\newcommand{\s}{\rm\; s}

\newcommand{\erg}{\rm\; erg}

\newcommand{\ergps}{\hbox{$\erg\s^{-1}\,$}}

\newcommand{\kmps}{\hbox{$\km\s^{-1}\,$}}

\newcommand{\kmpspMpc}{\hbox{$\kmps\Mpc^{-1}\,$}}

\newcommand{\Lx}{\hbox{$\thinspace L_\mathrm{X}$}}
\newcommand{\LEdd}{\hbox{$\thinspace L_\mathrm{Edd}$}}

\newcommand{\Omm}{\hbox{$\rm\thinspace \Omega_{m}$}}
\newcommand{\OmL}{\hbox{$\rm\thinspace \Omega_{\Lambda}$}}

\title[UMBHs in BCGs]{On the hunt for ultramassive black holes in brightest cluster galaxies}
\author[J. Hlavacek-Larrondo et al.]{J. Hlavacek-Larrondo$^{1}$\thanks{E-mail: juliehl@ast.cam.ac.uk}, A. C. Fabian$^{1}$, A. C. Edge$^{2}$ and M. T. Hogan$^{2}$\\
$^{1}$Institute of Astronomy, University of Cambridge, Madingley Road, Cambridge CB3 0HA\\
$^{2}$Institute of Computational Cosmology, Department of Physics, Durham University, Durham, DH1 3LE\\}
\begin{document}

\date{Accepted 2012 April 25. Received 2012 April 23; in original form 2011 October 25}

\pagerange{\pageref{firstpage}--\pageref{lastpage}} \pubyear{2011}

\maketitle

\begin{abstract}
We investigate where brightest cluster galaxies (BCGs) sit on the fundamental plane of black hole (BH) activity, an established relation between the X-ray luminosity, the radio luminosity and the mass of a BH. Our sample mostly consists of BCGs that lie at the centres of massive, strong cooling flow clusters, therefore requiring extreme mechanical feedback from their central active galactic nucleus (AGN) to offset cooling of the intracluster plasma ($L_{\rm mech}>10^{44-45}\ergps$). Based on the BH masses derived from the $M_{\rm BH}-\sigma$ and $M_{\rm BH}-M_{K}$ correlations, we find that all of our objects are offset from the plane such that they appear to be less massive than predicted from their X-ray and radio luminosities (to more than a 99 per cent confidence level). For these objects to be consistent with the fundamental plane, the $M_{\rm BH}-\sigma$ and $M_{\rm BH}-M_{K}$ correlations therefore seem to underestimate the BH masses of BCGs, on average by a factor of 10. Our results suggest that the standard relationships between BH mass and host galaxy properties no longer hold for these extreme galaxies. Furthermore, our results imply that if these BHs follow the fundamental plane, then many of those that lie in massive, strong cool core clusters must be ultramassive with $M_{\rm BH}>10^{10}{\rm M_\odot}$. This rivals the largest BH masses known and has important ramifications for our understanding of the formation and evolution of BHs.  
\end{abstract}


\section{Introduction}
Brightest cluster galaxies (BCGs) are the most massive galaxies in the present-day Universe. They lie at the very centres of galaxy clusters and exhibit some of the richest phenomena known. Many have an active galactic nucleus (AGN) at their core that is capable of inflating large cavities of relativistic plasma through jetted outflows. The energetics of these outflows often exceed $10^{44-45}\ergps$ in massive clusters and are sufficient, in the majority of clusters, to prevent catastrophic cooling of the surrounding hot X-ray emitting gas \citep{Fab2003344,Bir2004607,For2005635,Raf2006652,Fab2006366,Dun2006373,Dun2008385,San2007381,Mcn200745}. 

Although much is known about the properties of the outflows, the fundamental properties of the black holes (BHs) powering these outbursts remain largely unknown. Since BCGs are the most massive (and luminous) galaxies in the local Universe, by simple extrapolation of observed correlations, they should host the most massive BHs. Yet, only a handful have reliable BH mass measurements from dynamical modelling of the kinematics: M87 in Virgo (Macchetto et al. 1997), NGC1399 in Fornax (Houghton et al. 2006), NGC 6086 in A2162 (McConnell et al. 2011b), the BCGs in A3565, A1836 and A2052 (Dalla Bont\`a et al. 2009)\nocite{Mac1997489,Hou2006367,McC2011728,Dal2009690}, as well as NGC 3842 in A1367 and NGC 4889 in Coma (McConnell et al. 2011a)\nocite{McC2011480}. 

Furthermore, BCGs are galaxies that lie in some of the most extreme environments, subject to major mergers in the past, and powerful AGN feedback at present times. It is therefore not clear if they follow the standard $M_{\rm{BH}}-\sigma$ or $M_{\rm BH}-M_{K}$ correlations observed in the lower mass galaxies (see von der Linden et al. 2007, Dalla Bont\`a et al. 2009, and especially Lauer et al. 2007)\nocite{Lau2007662,von2007379,Dal2009690}. The few with reliable mass estimates suggest that some may follow the correlations, while others are offset such that the BH mass measured from dynamical modelling is larger than the values predicted from the $M_{\rm{BH}}-\sigma$ and $M_{\rm{BH}}-M_{K}$ correlations (e.g. M87, A1836-BCG, NGC 3842 and NGC 4889). McConnell et al. (2011a)\nocite{McC2011480} have also found the most massive BH to date (NGC 4889 with a $2.1\times10^{10}M_{\rm{\odot}}$ BH), providing the first direct evidence for the existence of ultramassive BHs (UMBHs, hereafter BHs with masses exceeding $10^{10}M_{\rm{\odot}}$).

The existence of UMBHs in BCGs has however already been predicted, especially for BCGs that lie in the most massive and strong cool core clusters ($\Lx>10^{45}\ergps$; $t_{\rm cool}<3$ Gyr). In these clusters, the central BH must be providing extreme amounts of energy to prevent the surrounding gas from cooling such that $L_{\rm mech}>10^{45}\ergps$. At these levels, the BH power exceeds $1$ per cent of the Eddington luminosity for a $10^9M_{\rm{\odot}}$ BH. \citet{Chu2005363} have argued, by analogy with the X-ray binaries, that BHs operating at such powers must be radiatively efficient, while at low powers (less than $1$ per cent of the Eddington luminosity), their radiative efficiency drops steadily and the power is increasingly taken up by outflows. This explains the behaviour of BHs in low/medium mass galaxy clusters, but cannot explain those in the most massive and strong cool core clusters with $>10^{45}$ erg/s jet powers. Since they are operating at powers exceeding $1$ per cent of the Eddington luminosity, they should be radiatively efficient (i.e. we should see an X-ray point source). However, in \citet{Hla2011} we showed that many of these extreme clusters showed no evidence of an X-ray point source at their centres. For them to be consistent with the lower mass clusters, we proposed that the BCGs host an UMBH at their centres ($>10^{10}{\rm M_\odot}$). In this case, the power only exceeds $0.1$ per cent of the Eddington luminosity and they do not require radiatively efficient nuclei. The existence of an UMBH has also been proposed to explain the unusually large pair of AGN-driven outflows in MS0735.6+7421 that require extreme jet powers of $10^{46}\ergps$ to create them \citep{McN2009698}. 

Here, we extend the work of \citet{Hla2011} and investigate where BCGs sit on the fundamental plane (FP) of BH activity. Our results show that the $M_{\rm BH}-\sigma$ and $M_{\rm BH}-M_{K}$ correlations for BCGs systematically underestimate the BH masses, and that many of the BCGs lying in massive and strong cool core clusters must have an UMBH at their centres with $M_{\rm BH}>10^{10}{\rm M_\odot}$. In Section 2, we present the sample and observations used to determine where the objects sit in the FP, and then in Section 3, we present the results. Finally, in Section 4 we discuss the results, and conclude in Section 5. We adopt $H_\mathrm{0}=70\kmpspMpc$ with $\Omm=0.3$, $\OmL=0.7$ throughout this paper. All errors are $2\sigma$ unless otherwise noted. 

\section{Fundamental Plane of black hole activity}

\subsection{Previous studies}
The FP of BH activity is an established correlation relating the mass of a compact object to its $2-10$ keV intrinsic X-ray  luminosity ($\Lx$) and $5$ GHz core radio luminosity ($L_{\rm 5GHz}$), see \citet[][]{Mer2003345,Fal2004414,Kor2006456,Gul2009706,Plo2012419} (hereafter MHD2003, FKM2004, KFC2006, GCM2009 and PMK2012 respectively). The FP covers over six orders of magnitude in mass and is especially important since it provides a unification scheme for all BHs and X-ray binaries (XRB), regardless of their mass. 

\begin{table*}
\addtolength{\tabcolsep}{-2pt}
\caption{Sample of Clusters - (1) Name; (2) redshift; (3) Core radio luminosity at 5 GHz. The 11 BCGs that have multi-frequency radio data available on them are indicated with the $^{S,F}$ symbols, for steep ($\alpha>0.4$) and flat/inverted ($-0.2<\alpha<0.4$)  spectrum sources respectively; (4) Intrinsic rest-frame X-ray luminosity in the $2-10$ keV band of the nucleus (note that all are upper limits, except for Cygnus A and M87); (5) 2MASS $K$ band absolute magnitude of the bulge; (6) $M_{\rm BH}$ estimated from the $K$ band bulge luminosity using Eq. \ref{eq2}; (7) $M_{\rm BH}$ if the object sat on the fundamental plane, applying an average correction factor of $\log{\Delta}M_{\rm BH}=0.8\pm0.6$ to the $K$ band derived BH mass; (8) Eddington ratio of the nuclear X-luminosity, using the BH mass in Column 6; (9) Eddington ratio of the nuclear X-luminosity, using the BH mass in Column 7. (10) Notes concerning the radio luminosities: i) derived from VLA observations at 4.9 GHz; ii) derived from ATCA observations at 5.5 GHz; iii) extrapolated from the core 1.4 GHz flux and core spectral index \citep[$\alpha=0.56$;][]{Gia2011525}; iv) no detection at 4.9 GHz; v) private communications (A. Edge); vi) extrapolated from the 1.4 GHz, 8.46 GHz and 28.5 GHz (BIMA) flux densities; vii) determined from EMSS; viii) from MH2007. All errors are 2$\sigma$. $^aK$ band absolute magnitude derived from UKIRT.}
\begin{tabular}{lccccccccc}
\hline
\hline
(1) & (2) & (3) & (4) & (5) & (6) & (7) & (8) & (9) & (10) \\
Cluster & $z$ & log$L_{\rm 5 GHz}$ &  log$L_{\rm X}$ & $M_{K}$ & log$M_{\rm BH,}{_K}$ & log$M_{\rm BH, FP}$ & log$\lambda_{M_{\rm BH,}{_K}}$ & log$\lambda_{M_{\rm BH, FP}}$ & Note \\
\hline
A1835 & 0.2532 & 41.07$\pm$0.02$^S$ & $<$42.25$\pm$0.19 & -27.36$\pm$0.28 & 9.5$\pm$0.4 & 10.3$\pm$0.7 & -5.4$\pm$0.4 & -6.2$\pm$0.7 & (i)\\
A2204 & 0.1522 & 40.81$\pm$0.02 & $<$42.21$\pm$0.13 & -26.57$\pm$0.27$^a$ & 9.2$\pm$0.3 & 10.0$\pm$0.6 & -5.1$\pm$0.3 & -5.9$\pm$0.7 & (i)\\
A1664 & 0.1283 & 40.51$\pm$0.02$^F$ & $<$41.18$\pm$0.48 & -26.15$\pm$0.22 & 9.1$\pm$0.3 & 9.9$\pm$0.7 & -6.0$\pm$0.5  & -6.8$\pm$0.8 & (ii)\\
RXCJ1504.1-0248 & 0.2153 & 41.15$\pm$0.05$^F$ & $<$42.43$\pm$0.28 & -26.59$\pm$0.36 & 9.2$\pm$0.3 & 10.0$\pm$0.7 & -4.9$\pm$0.4 & -5.7$\pm$0.7 & (iii)\\
RXJ0439.0+0715 & 0.2300 & $<$39.18 & $<$41.76$\pm$0.51 & -26.31$\pm$0.36 & 9.1$\pm$0.3 & 9.9$\pm$0.7 & -5.5$\pm$0.6 & -6.3$\pm$0.8 & (iv)\\
A2390 & 0.2280 & 41.95$\pm$0.02 & $<$42.03$\pm$0.20 & -27.05$\pm$0.34 & 9.4$\pm$0.3 & 10.2$\pm$0.7 & -5.5$\pm$0.4 & -6.3$\pm$0.7 & (v)\\
A0478 & 0.0881 & 39.85$\pm$0.03$^S$ & $<$41.53$\pm$0.20 & -26.72$\pm$0.14 & 9.3$\pm$0.3 & 10.1$\pm$0.7 & -5.9$\pm$0.3 & -6.7$\pm$0.7 & (vi)\\
PKS0745-19 & 0.1028 & 40.13$\pm$0.02$^S$ & $<$41.54$\pm$0.55 & -26.87$\pm$0.18 & 9.4$\pm$0.3 & 10.1$\pm$0.7 & -5.9$\pm$0.6 & -6.7$\pm$0.9 & (v)\\
A2261 &  0.2240 & 39.70$\pm$0.07$^S$ & $<$41.58$\pm$0.73 & -27.35$\pm$0.20 & 9.5$\pm$0.3 & 10.3$\pm$0.7 & -6.0$\pm$0.8 & -6.8$\pm$1.0 & (i)\\
Z2701 & 0.2151 & 40.40$\pm$0.02$^F$ & $<$41.61$\pm$0.47 & -26.26$\pm$0.34 & 9.1$\pm$0.3 & 9.9$\pm$0.7 & -5.6$\pm$0.5 & -6.4$\pm$0.8 & (i)\\
RXJ1720.1+2638 & 0.1640  & 40.00$\pm$0.04$^F$ & $<$41.90$\pm$0.35 & -26.71$\pm$0.22 & 9.3$\pm$0.3 & 10.1$\pm$0.7 & -5.5$\pm$0.4 & -6.3$\pm$0.8 & (i) \\
RXJ2129.6+0005 & 0.2350 & 40.81$\pm$0.02$^S$ & $<$42.23$\pm$0.27 & -26.73$\pm$0.30 & 9.3$\pm$0.3 & 10.1$\pm$0.7 & -5.2$\pm$0.4 & -6.0$\pm$0.7 & (i)\\
Z3146 & 0.2906 & 40.02$\pm$0.03$^S$ & $<$42.85$\pm$0.13 & -26.45$\pm$0.56 & 9.2$\pm$0.3 & 10.0$\pm$0.7 & -4.4$\pm$0.3 & -5.2$\pm$0.7 & (i)\\
MS1455.0+2232 & 0.2578 & 40.34$\pm$0.05$^F$ & $<$42.29$\pm$0.19 & -27.13$\pm$0.28 & 9.4$\pm$0.3 & 10.2$\pm$0.7 & -5.3$\pm$0.4 & -6.1$\pm$0.7 & (i)\\
MS2137.3-2353 & 0.3130 & 40.20$\pm$0.02 & $<$42.88$\pm$0.26 & -26.77$\pm$0.30 & 9.3$\pm$0.3 & 10.1$\pm$0.7 & -4.5$\pm$0.4 & -5.3$\pm$0.7 & (vii)\\
Centaurus & 0.0104 & 39.10$\pm$0.02 & $<$39.41$\pm$0.12 & -26.14$\pm$0.04 & 9.1$\pm$0.2 & 9.9$\pm$0.6 & -7.8$\pm$0.3 & -8.6$\pm$0.7 & (viii)\\ 
Cygnus A & 0.0561 & 41.43$\pm$0.02 & 44.36$\pm$0.02 & -26.73$\pm$0.12 & 9.3$\pm$0.3 & 10.1$\pm$0.7 & -3.1$\pm$0.3 & -3.9$\pm$0.7 & (viii) \\
M87 &  0.00436 & 38.88$\pm$0.02 &  40.55$\pm$0.04 & -25.55$\pm$0.04 & 8.8$\pm$0.2 & 9.6$\pm$0.6 & -6.4$\pm$0.2 & -7.2$\pm$0.6 & (viii) \\
\hline
\end{tabular}
\label{tab1}
\end{table*}

MHD2003 compiled the first detailed study of the FP based on a large sample comprising around 100 AGNs and 8 galactic BHs. The sample consisted of a diverse population of objects, including X-ray Binaries (XRB), low-luminosity AGN (LLAGN), low ionization nuclear emission region (LINER), type 1 and 2 Seyfert galaxies, Fanaroff-Riley (FR) radio galaxies, as well as radio loud and radio quiet quasars. Only direct measurements of $L_{\rm 5GHz}$ and $L_{\rm X}$ were used. This study therefore finds the existing correlation of the FP averaged over all types of BHs, regardless of the accretion state. The best-fitting relationship found by the authors is given in Eq. \ref{eq1}. 
\begin{equation}
{\log}L_{\rm 5GHz}=0.60{\log}L_{\rm X} + 0.78{\log}M_{\rm BH} + 7.33
\label{eq1}
\end{equation}

On the other hand, FKM2004 argued that the FP is due to synchrotron emission arising from a jetted BH, and therefore only BHs in the ``low/hard" (LH) state should be included in the analysis since this state is dominated by jet emission \citep{Fen2001322}. More recently, both the studies by MHD2003 and FKM2004 have been revised by KFC2006, with the aim of improving the parameter estimates of the FP. Their study suggested that the resulting parameters depend strongly on the weights given to each AGN class, and on the assumptions made on the sources of scatter (e.g. relativistic beaming and non-simultaneous measurements of X-ray/radio luminosities). They also found that sub-Eddington sources equivalent to the LH state seem to follow the plane more tightly. In an effort to refine the scatter in the FP, GCM2009 only considered sources where dynamical mass-measurements were available (18 objects in total). Although this limited the sample to nearby AGN ($<30$ Mpc), the authors found that when the sample was further limited to low accreting sources (from ${\log}{\lambda}={\log}({\Lx}/{\LEdd})={-4.2}$ to ${\log}{\lambda}=-5.2$), the scatter in the best-fitting relation decreased, suggesting once more that high accreting sources may not belong to the FP. The majority of BCGs we consider here, are those that lie in massive, strong cool core clusters with powerful outflows ($L_{\rm {mech}}>10^{44-45}\ergps$) and radiatively inefficient nuclei ($\Lx~{\aplt}~10^{42}\ergps$). The majority are therefore highly sub-Eddingtion (${\log}{\lambda}~\aplt~{-5}$, see Table \ref{tab1}), and are considered to be in a state equivalent to the LH state.

PMK2012 have used a more sophisticated regression technique (Bayesian), allowing them to further constrain the FP and obtain smaller uncertainties on their best-fitting coefficients. They use a sample of low-accreting BHs and find that their best-fitting parameters favour the coefficients predicted for X-rays that are dominated by optically thin jet emission, and not coronal emission, if their objects have flat/inverted radio emission. However, they also find that the most massive BHs ($\apgt10^8M_{\rm \odot}$) could be strongly affected by synchrotron cooling (see also FKM2004 and KFC2006). We discuss this issue further in Section 3.2.       

Our aim is to determine where BCGs sit on the FP of BH activity for the general population of compact objects. We therefore initially adopt the original correlation from MHD2003 to illustrate our results, since it includes all sources regardless of the state in which they are in. However, the best-fitting relations found by the more recent studies of KFC2006, GCM2009 and PMK2012 are also included in our analysis of Section 3, and we show that our results are still consistent when using these correlations. 

\begin{figure*}
\centering
\begin{minipage}[c]{0.99\linewidth}
\centering \includegraphics[width=\linewidth]{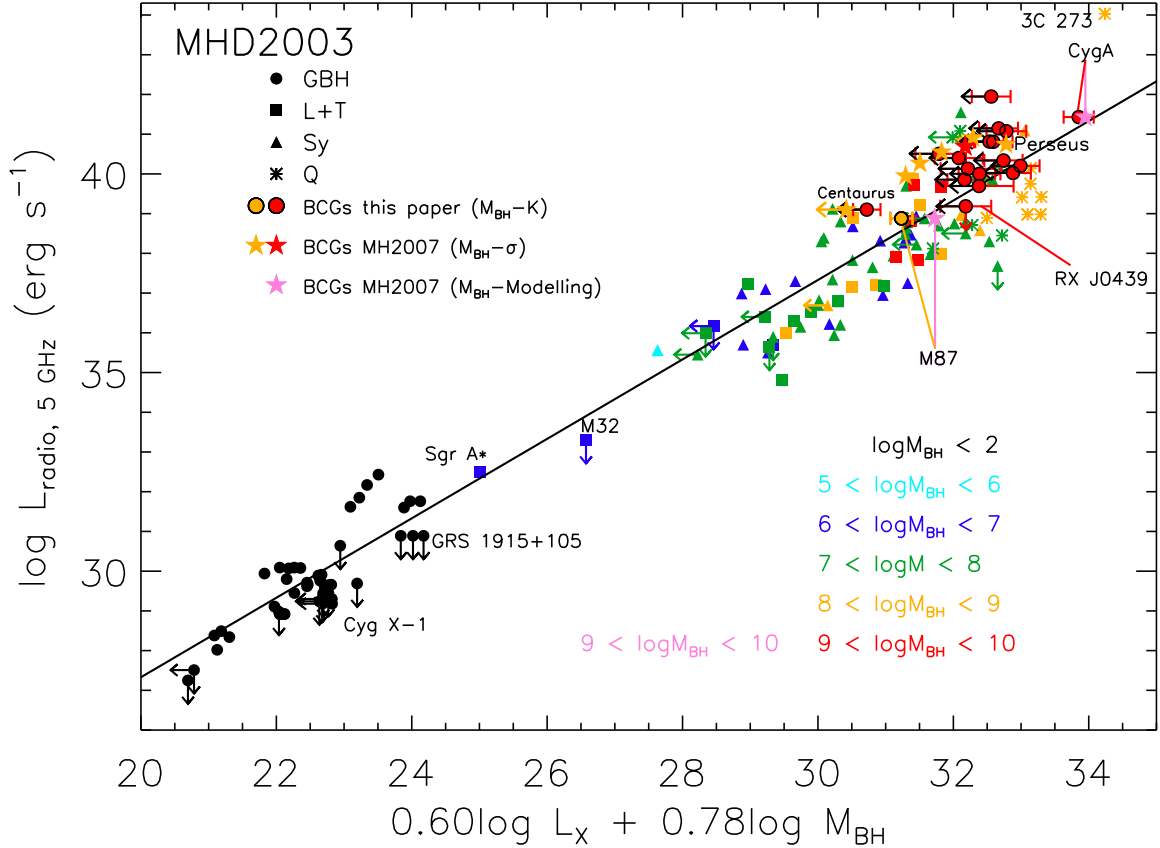}
\end{minipage}
\caption{Fundamental Plane of BH activity. We adopt the relation derived by MHD2003 and show the best-fitting regression with the solid line (see Eq. \ref{eq1}). The sample of MHD2003 contains Galactic BHs (GBH), LINERs (L), transition nuclei (LINER/H\thinspace{\sc ii}; T), Seyfert galaxies (Sy) and quasars (Q). We include the 18 BCGs from Table \ref{tab1} (shown with the filled red and yellow circles), 16 of which have no detectable X-ray nucleus. For all of these objects, the BH masses were derived using the $M_{\rm BH}-M_{K}$ relation (Eq. \ref{eq2}). We also include the 9 BCGs from MH2007, 7 of which have BH masses derived from the $M_{\rm BH}-\sigma$ relation (shown with the 5 pointed red and yellow stars). The remaining two (M87 and Cygnus A) have BH masses derived form dynamical modelling and are shown with the pink stars. Our results show that BCGs lie on or above the relation, i.e. are offset from the plane such that they appear too bright in the radio compared to their X-ray luminosities and predicted BH masses. A possible explanation is that the $M_{\rm BH}-\sigma$ and $M_{\rm BH}-M_{K}$ correlations systematically underestimate the masses of these BHs, implying that if these objects follow the FP, some would be ultramassive ($M_{\rm BH}>10^{10}{\rm M_\odot}$).  }
\label{fig1}
\end{figure*}

\subsection{Brightest Cluster Galaxies}
Our initial sample of BCGs include the 19 clusters studied in \citet{Hla2011} and consists of 18 clusters that have no detectable central X-ray nucleus, as well as Cygnus A which has clear non-thermal nuclear emission in the form of an absorbed power law. We use the values obtained in \citet{Hla2011} for $\Lx$, all of which were derived from $Chandra$ X-ray observations. To obtain the values, we converted background subtracted number of counts of the nuclear region, within a $1''$ radius equivalent to the Chandra point spread function, into fluxes using the web interface PIMMS\footnote[1]{http://heasarc.gsfc.nasa.gov/Tools/w3pimms.html} \citep[][]{Mukai1993}. Here, we include Galactic absorption and model the emission in the form of a power law with a spectral index equal to 1.9. Since the counts are dominated by thermal emission, with no clear evidence of non-thermal emission in the spectra, the fluxes we obtain are considered to be upper limits to the non-thermal contribution of the nucleus. The corresponding luminosities, along with the $2\sigma$ errors are shown in Column 4 of Table \ref{tab1}.

The 5 GHz radio luminosities are obtained using a combination of published values and archival maps, all of which have beam sizes $<5''$ (high angular resolution is needed to isolate the core emission). For RXC J1504.1-0248, $L_{\rm 5GHz}$ was derived using the 1.4 GHz core flux measurement ($S_{\rm 1.4GHz}=42.2$ mJy) and the core spectral index ($\alpha=0.56$; $S_\nu\propto\nu^{-{\alpha}}$) of \citet{Gia2011525}. For A0478, we extrapolated the 5 GHz flux based on the peak Very Large Array (VLA) 1.4 GHz and 8.46 GHz fluxes (A configuration, project code AE117), as well as the 28.5 GHz measurement from the Berkeley-Illinois-Maryland Association \citep[BIMA;][]{Cob2008134}. The 5 GHz flux from the Einstein Observatory Extended Medium-Sensitivity Survey \citep[EMSS;][]{Gio199072,Sto199176} was used for MS2137.3-2353. Finally, we adopted the 5 GHz core radio luminosities from \citet[][ hereafter MH2007]{Mer2007381} for the Centaurus cluster and Cygnus A. For the remaining objects, we computed the 5 GHz radio luminosities from the 4.9 GHz C configuration VLA observations (project codes AE099, AE107, AE125; PI Edge) or 5.5 GHz Australia Telescope Compact Array (ATCA) observations (project code C1958), and use a conservative estimate by taking the peak intensity as the 5 GHz flux measurement. Our results are shown in Column 3 of Table \ref{tab1}. Errors are derived as the quadratic sum of the rms noise level in the radio map and the systematic error associated with the value. Systematic errors vary with frequency, but are on the order of 5 per cent \citep[see][]{Car1991383}. For simplicity, we therefore choose to compute the total error assuming a 5 per cent systematic error and a 2$\sigma_{\rm rms}$ noise level. For the values taken from the literature where no error estimate was available, we only considered systematic uncertainties and assume a 5 per cent uncertainty on the value.  

We then calculate the predicted BH masses for our BCGs using the $M_{\rm BH}-M_{K}$ relation recently revised in detail by \citet[][]{Gra2007379} and shown in Eq. \ref{eq2}. 
\begin{equation}
\resizebox{.90\hsize}{!}{${\log}(M_{\rm BH}/{M_\odot})=-0.37^{{\pm}0.04}(M_{K}+24) + 8.29^{{\pm}0.08}$}
\label{eq2}
\end{equation}

We use $K$ band Two Micron All Sky Survey (2MASS) magnitudes from the All-Sky Extended Source Catalogue determined from fit extrapolation \citep[][]{Jar2000119}, which we correct for Galactic extinction \citep{Sch1998500}. We also correct for redshift ($K$-correction) following the latest {\sc IDL} scripts\footnote[2]{http://howdy.physics.nyu.edu/index.php/Kcorrect} from \citet[][]{Bla2007133} and for evolution ($E$-correction) using the $K$ band 2MASS correction estimated in \citet[][ $0.8\times{z}$]{Bel2003149}. The resulting values and 2$\sigma$ errors are shown in Column 5 of Table \ref{tab1}. 

Three clusters did not have 2MASS observations (A2204, MACS J1532.8+3021 and RX J1347.5-1145), but A2204 has a published United Kingdom Infra-Red Telescope (UKIRT) absolute $K$ band magnitude \citep{Sto2008384}, where the authors estimate that the dominant error in their photometry is due to the fitting algorithm which can underestimate the integrated brightness by up to 10 per cent of a magnitude. The error we show in Table \ref{tab1} for A2204 is therefore taken as 10 per cent of the value. Since the remaining two clusters (MACS J1532.8+3021 and RX J1347.5-1145) do not have 2MASS/UKIRT magnitudes, we do not include them in our final sample.  

The BH masses are then estimated using the absolute magnitudes ($M_{K}$) and Eq. \ref{eq2}. We use a Monte Carlo technique to calculate the errors in the derived masses, where we assume that both parameters in Eq. \ref{eq2} (0.37$\pm$0.04 and 8.29$\pm0.08$), as well as $M_{K}{\pm}{\Delta}M_{K}$ are independent from one another. We also assume that each follow a Gaussian distribution based on their values and associated errors. For each object, we proceed by selecting 100 random variables for all three distributions (0.37$\pm$0.04, 8.29$\pm0.08$ and $M_{K}{\pm}{\Delta}M_{K}$), and run through all possibilities, each time computing the predicted BH mass. The final BH masses are then taken as the median values, and the $2\sigma$ errors are calculated within the 2.2 and 97.6 percentiles (see Column 6 of Table \ref{tab1}). 

Finally, we plot the location of each BCG on the FP following Eq. \ref{eq1} in Fig. \ref{fig1}. Fig. \ref{fig1} also includes the 9 BCGs from MH2007: Cygnus A, NGC 1275 in Perseus, NGC 4486 in M87, NGC 4696 in Centaurus, NGC 6166 in A2199, IC 4374 in A3581, UGC 9799 in A2052, 3C 218 in Hydra A and 3C 388. In MH2007, the BH masses are derived from the $M_{\rm BH}-\sigma$ relation or from dynamical mass measurements for M87 and Cygnus A. Note that Table \ref{tab1} already includes NGC 4696 (Centaurus) and Cygnus A, and shows the BH masses predicted from the $M_{\rm BH}-M_{K}$ relation. We also include M87 in Table \ref{tab1} and derive its BH mass as predicted from the $M_{\rm BH}-M_{K}$ relation (from its 2MASS $K$ band magnitude), in order to compare the predicted mass to the one measured from modelling of the kinematics. The X-ray and radio luminosities for M87 were taken from MH2007, where we assume a 5 per cent uncertainty on the first (equivalent to the systematic uncertainty in radio) and a 10 per cent uncertainty on the second. Finally, we include all the data points from MHD2003 in Fig. \ref{fig1}, but remove Cygnus A, NGC 4486 (M87), NGC 1275 (Perseus) and NGC 6166 (A2199), since they are already included in MH2007. 

In Fig. \ref{fig2}, we only show the location of the BCGs in the FP to illustrate more clearly where BCGs lie on the FP. Note that, most of our BCGs have no detectable X-ray nucleus (shown with the arrows), implying that they could lie even more leftward than shown in Fig. \ref{fig2}.

\section{Results}

\begin{figure}
\begin{minipage}[c]{0.99\linewidth}\hspace{-0.6cm}
\centering \includegraphics[width=\linewidth]{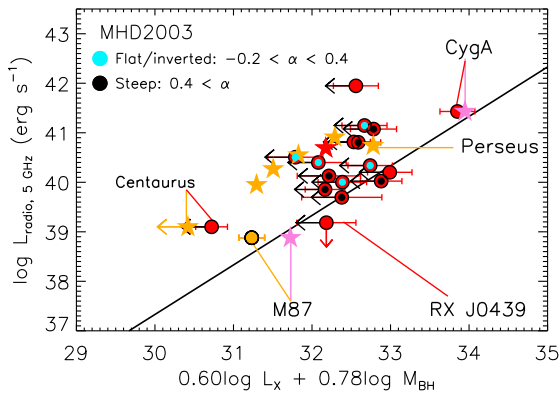}
\end{minipage}
\caption{Fundamental Plane of BH activity. Same as Fig. \ref{fig1}, but focussed on the BCGs. Also highlighted are the 11 BCGs that have multi-frequency radio data available on them with the steep spectrum sources ($\alpha>0.4$) shown in black and the flat/inverted spectrum sources ($-0.2<\alpha<0.4$) shown in light blue.}
\label{fig2}
\end{figure}

Figs. \ref{fig1} and \ref{fig2} show that our BCGs systematically sit on or above the FP of BH activity. The only outlier is RX~J0439.0+0715 which has no detectable X-ray or radio nucleus. The BCGs are offset from the plane such that they appear too bright in the radio compared to their X-ray luminosities and predicted BH masses. Most of our objects lie in massive, strong cool core clusters that have powerful outflows ($L_{\rm {mech}}>10^{44-45}\ergps$) and no detectable X-ray nucleus ($\Lx~{\aplt}~10^{42}\ergps$). The majority of our objects are highly sub-Eddingtion with ${\log}{\lambda}~\aplt~{-5}$ (except for Cygnus A, see Table \ref{tab1}). We therefore consider them to be in a state equivalent to the LH state of X-ray binaries. 

Although the scatter of the FP is large, it is unlikely that all BCGs lie randomly above the best-fitting relation; they should on average lie above and below the plane (i.e. not systematically above). We therefore examine four different possibilities as to why BCGs appear to be systematically offset from the plane. 

\subsection{Overestimating radio luminosities}

First, we investigate the possibility that we could be overestimating the radio luminosities of all our BCGs. 

It is not clear why this would only be the case for BCGs considering that other AGNs such as Seyferts and quasars are also part of the MHD2003 sample and do not seem to be affected by such a bias. The 5 GHz radio luminosities of the MHD2003 sample were obtained from the literature. Their sample consists both of steep ($\alpha>0.4$) and flat ($\alpha<0.4$) spectrum sources, and most of the luminosities were obtained from arcsec resolution VLA measurements and by integrating the fluxes. We have used a conservative approach to our 5 GHz luminosities, based on the peak fluxes and not integrated fluxes (peak fluxes are smaller than integrated ones). We should therefore not be overestimating the radio luminosities compared to the objects in MHD2003, yet our objects seem to lie above the FP.

PMK2012 showed that the FP slopes change when steep spectrum sources are included. They also derive what they consider to be the most accurate FP regression to date (with the least scatter), based only on sub-Eddington accreting BHs with flat/inverted radio spectra, thus suggesting that only these sources should be included in the FP. Eleven of our objects in Table \ref{tab1} have multi-frequency radio data available on them (mostly from the VLA archive), 6 of which have steep spectra ($\alpha>0.4$) and are shown in Column 3 of Table \ref{tab1} with the $^S$ symbol. The remaining 5 have flat spectra with $\alpha$ varying between $-0.2$ to $+0.4$ and are shown in Column 3 of Table \ref{tab1} with the $^F$ symbol. Both steep and flat spectrum sources are also shown in Fig. \ref{fig2} with the black and blue points, respectively. If the offset observed for BCGs in the FP were due to the wrongfully included steep spectrum sources, we would have expected these sources to systematically have the largest offsets. However, Fig. \ref{fig2} shows that this is not the case, and that both steep and flat spectrum seem to spread out randomly in the scatter. It therefore seems unlikely that PMK2012's interpretation would explain why our objects are systematically offset from the plane.

\subsection{Underestimating X-ray luminosities: synchrotron cooling}

FKM2004 and KFC2006 argued that the X-ray emission for the most massive BHs ($\apgt10^8M_{\rm {\odot}}$) could be strongly affected by synchrotron cooling. Synchroton cooling is anti-correlated with BH mass, and for the most massive BHs, the cooling break occurs below X-ray wavelengths. Synchrotron cooling alters the spectral distribution above the cooling break and causes the X-ray luminosities to be underestimated from the true value for the most massive BHs. This makes BHs appear underluminous in the X-rays compared to their radio luminosities. We expect our objects to host at least $10^{8-9}M_{\rm {\odot}}$ BHs. Synchrotron cooling could therefore explain why our objects appear to be so radiatively inefficient and why they appear to be offset from the plane. 

PMK2012 have explored this issue in more detail, and determined that even optical emission can be affected by synchrotron cooling in the most massive objects such as FR~I galaxies. They argue that spectral energy distribution (SED) modelling is necessary to obtain the luminosities and place the objects on the FP. Although this could explain why our objects are offset from the plane, SED modelling is beyond the scope of this paper, and we examine other possibilities as to why BCGs appear to lie above the FP of BH activity.

\subsection{Exceptional BHs in BCGs}

The third possibility is that BCGs are special in the sense that they occupy a particular place in the FP. This would imply that the BHs in these systems are operating in a different way than other BHs, which in itself is interesting. BCGs lie at the very centres of galaxy clusters and are surrounded by a substantial amount of hot dense gas. The most noticeable difference between BCGs and systems such as quasars and Seyferts is therefore the environment. It could be that the dense hot gas provides a unique environment that makes BCGs intrinsically more radio luminous for a given X-ray luminosity and BH mass, if the radio emission is due to synchrotron emission.

\subsection{Underestimating BH masses from the $M_{\rm BH}-\sigma$ and $M_{\rm BH}-M_{K}$ relations}

Finally, we examine the possibility that the $M_{\rm BH}-\sigma$ and $M_{\rm BH}-M_{K}$ correlations systematically underestimate the BH masses in BCGs. We base this idea on the recent studies that have obtained direct BH masses from dynamical modelling of the kinematics for a handful of BCGs (Macchetto et al. 1997; Houghton et al. 2006; Dalla Bont\`a et al. 2009; McConnell et al. 2011a,b)\nocite{Mac1997489,Hou2006367,McC2011728,Dal2009690,McC2011480}, many of which seem to indicate that direct BH mass measurements from dynamical modelling are larger than the values predicted from the $M_{\rm{BH}}-\sigma$ and $M_{\rm{BH}}-M_{K}$ correlations (e.g. M87, A1836-BCG, NGC 3842 and NGC 4889). 

\begin{table}
\addtolength{\tabcolsep}{-4pt}
\caption{Average BH mass offset from the FP - (1) Offset for the 18 BCGs that have a mass estimate from the $M_{\rm BH}-M_{K}$ correlation (Table \ref{tab1}). The errors are 2$\sigma$ and were determined using a Monte Carlo technique, see Section 3.4. (2) Probability that the mass offset is larger than zero in terms of per cent (and $\sigma$). (3) Rough calculation of the average offset including the 18 BCGs in Table \ref{tab1}, as well as the 7 BCGs from MH2007 that have a mass estimate from the $M_{\rm BH}-\sigma$ correlation. }
\begin{tabular}{lccc}
\hline
\hline
& (1) & (2) & (3) \\
& $\log{\Delta}M_{\rm BH,}{_K}$ & Prob($>0$) & $\log{\Delta}M_{\rm BH,}{_{K-\sigma}}$\\
\hline
MHD2003 & 0.8$\pm$0.6 & $>99.6$ ($\apgt3\sigma$) & 1.0\\
KFC2006~[MHD] & 1.5$\pm$0.7 & $>99.9$ ($>3\sigma$) & 1.6 \\
\hspace{1.35cm}[FKM] & 2.1$\pm$0.7 & $>99.9$ ($>3\sigma$) & 2.3\\
GCM2009 & 0.5$\pm$0.6 & $>94.6$ ($\apgt2\sigma$) & 0.7 \\
PMK2012 & 2.6$\pm$0.7 & $>99.9$ ($>3\sigma$) & 2.8 \\
\hline
\end{tabular}
\label{tab2}
\end{table}
We do not expect all BCGs to intersect exactly the plane. However, it is unlikely that they randomly all lie above the plane in Fig. \ref{fig2}. Instead, they should on average lie above and below the plane.

To illustrated this, we determine the average mass offset needed ($\log{\Delta}M_{\rm BH}$) for BCGs to be consistent with the FP such that on average, they satisfy Eq. \ref{eq3}. $\log{\Delta}M_{\rm BH}$ therefore represents the offset needed so that BCGs lie on \emph{average} above and below the plane. 
\begin{equation}
\resizebox{.90\hsize}{!}{${\log}L_{\rm 5GHz}=0.60{\log}L_{\rm X} + 0.78{\log}(M_{\rm BH_{K,\sigma}}+{\Delta}M_{\rm BH}) + 7.33$~\label{eq3}}
\end{equation}

All the clusters in Table \ref{tab1} have error measurements for $L_{\rm 5GHz}$, $\Lx$ and $M_{\rm {BH,}}{_K}$ (from $K$ band magnitudes), thus allowing us to determine the average mass offset needed by using a Monte Carlo technique. Here, we assume that $L_{\rm 5GHz}$, $\Lx$ and $M_{\rm {BH,}}{_K}$ are independent, and that each follow a Gaussian distribution based on their values and associated uncertainties. For each of the 18 BCGs in Table \ref{tab1}, we assign 500 random variables to the $L_{\rm 5GHz}$ distribution, 500 to the $\Lx$ distribution and 500 to the $M_{\rm {BH,}}{_K}$ distribution. Using 1000 variables instead of 500 for each distribution does not change our results. Then, for each of the $500^3$ possibilities and for each of the 18 objects, we calculate the $\log{\Delta}M_{\rm {BH,}}{_K}$ needed for the object to satisfy Eq. \ref{eq3}. For each of the $500^3$ possibilities, we then calculate the average value of $\log{\Delta}M_{\rm {BH,}}{_K}$ over the 18 objects. This allows us to build a distribution containing $500^3$ estimates of the average offset needed for BCGs to be consistent with the FP. Our results do not change significantly if median values are used instead of average values. 

The final mass offset is taken as the median value in the $\log{\Delta}M_{\rm {BH,}}{_K}$ distribution and the $2\sigma$ errors are taken within the 2.2 and 97.6 percentiles. We repeat this calculation for 5 different FP regressions: MHD2003, both for the revised relations of FKM2004 and MHD2003 in KFC2006, GCM2009 and PMK2012. The results are shown in Column 1 of Table \ref{tab2}. We emphasize that 16 of the 18 BCGs in Table \ref{tab1} have no detectable X-ray nucleus. If we were to remove the 2 BCGs with X-ray nuclei from the calculations, we would obtain the same mass offsets. The calculations are therefore dominated by the non-detections, and the mass offsets should be regarded as the minimum offset needed for BCGs to lie on the FP. 

We have also considered the 7 BCGs from MH2007, which have BH mass measurements based on the $M_{\rm BH}-\sigma$ relation. Since MH2007 do not include uncertainty measurements on their values, we only compute the average mass offset needed for all BCGs to be consistent with the FP such that they satisfy Eq. \ref{eq3} (no Monte Carlo calculations performed over the uncertainties). Here, all 18 BCGs from Table \ref{tab1} and the 7 BCGs from MH2007 are included in the calculation. Note that NGC 4696 (Centaurus) is counted twice in the calculation, once for the mass derived from the $K$ band magnitude and once for the mass derived from the velocity dispersion ($\sigma$). The offsets we find are shown in Column 3 of Table \ref{tab2} and represent the offsets needed such that all 25 BCGs lie on average above and below the best-fitting plane. 

\begin{figure*}
\centering
\begin{minipage}[c]{0.99\linewidth}
\centering \includegraphics[width=\linewidth]{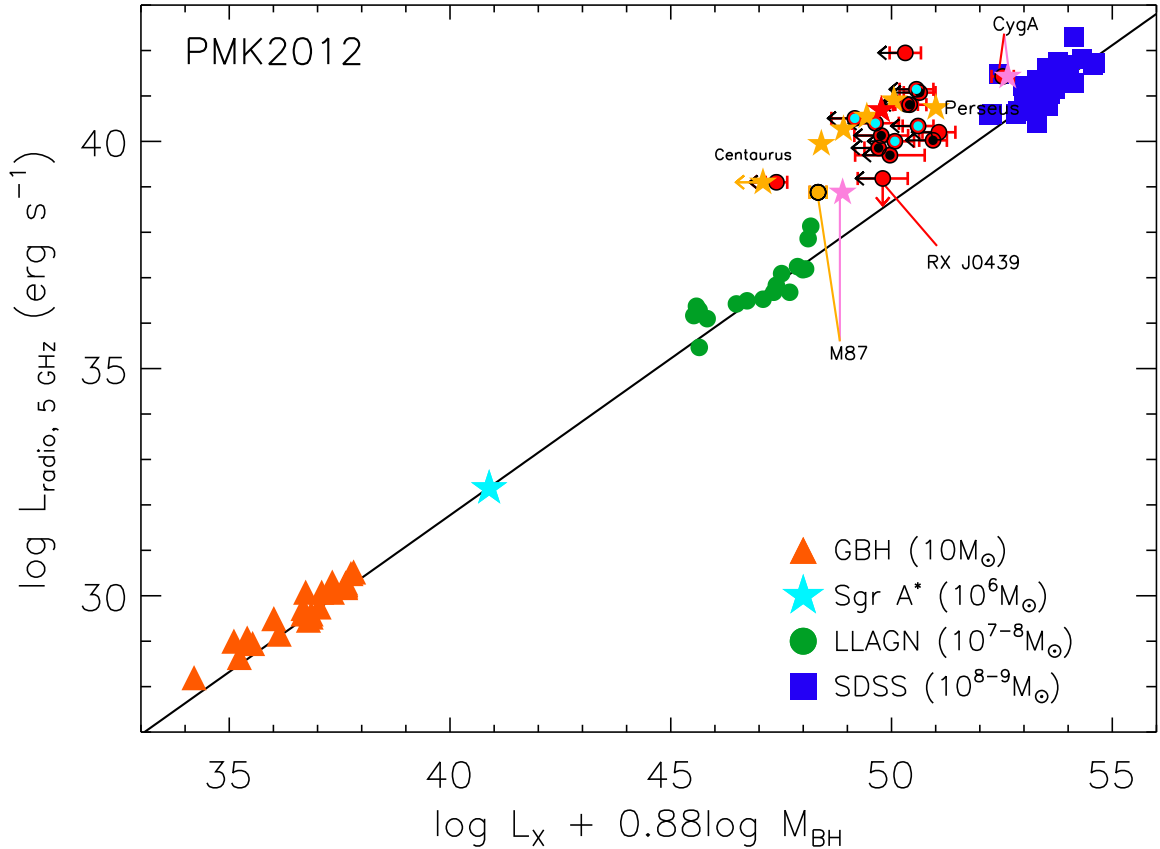}
\end{minipage}
\caption{Same as Fig. \ref{fig1}, but where we consider the best-fitting relation of PMK2012 (black line) which is considered to be the most robust to date. It is also the relation that has the largest mass offset for BCGs from the FP (see Table \ref{tab2}). Shown are the data points used in PMK2012 to derive the best-fitting relation, which only include sub-Eddington BHs with flat/inverted radio spectra (GBHs, Sgr A$^*$, LLAGN, as well as HBLs from SDSS). The BCGs are illustrated with the same symbols as in Fig. \ref{fig2} and are offset from the plane such that they appear to be less massive than predicted from their X-ray and radio luminosities. A possible explanation is that the $M_{\rm BH}-\sigma$ and $M_{\rm BH}-M_{K}$ correlations systematically underestimate the masses of BHs in BCGs. }
\label{fig3}
\end{figure*}
Table \ref{tab2} shows that all of the offsets are positive, i.e. consistent with the idea that the $M_{\rm BH}-\sigma$ and $M_{\rm BH}-M_{K}$ correlations systematically underestimate BH masses in BCGs. Interestingly, we find that the PMK2012 regression, considered to be the most robust to date, has the largest mass offset. Since this is currently the most accurate FP regression, we show the location of our BCGs in this plane in Fig. \ref{fig3}. Note that, PMK2012 only considered sub-Eddington BHs (i.e. BHs considered to be in the LH state) with flat/inverted radio spectra. By limiting the sample to these sources, they obtained the most tightly constrained regression coefficients to date. Their sample consists of GBHs, Sgr A$^*$, LLAGN and 39 BL Lac objects from the Sloan Digital Sky Survey (SDSS). The latter are relativistically beamed sub-Eddington AGN where the jet core emission can be better isolated due to the beaming. The sample of BL Lac objects only includes those that have peak synchrotron luminosities in the soft X-rays, referred to as high-energy cutoff BL Lac objects (HBL). In this case, the X-ray emission is mostly considered to be synchrotron emission, whereas with low-energy cutoff BL Lac objects (LBL), the peak synchrotron luminosities occur at near-infrared wavelengths, and the synchrotron emission is most likely contaminated with synchrotron self-Compton and external inverse Compton emission. Since the FP is thought to arise from BHs in the LH state dominated by synchrotron jet emission, it is necessary to only consider the sample of HBL as opposed to LBL. In Fig. \ref{fig3} we show PMK2012's best-fitting relation along with their sample of 82 objects. This figure shows clearly that our BCGs lie significantly offset from the plane, such that they appear to be less massive than that predicted from their X-ray and radio luminosities.  
   
Table \ref{tab2} also shows that the best-fitting relation by GCM2009 is only consistent with a positive offset to a $\sim95$ per cent confidence level. GCM2009 only uses objects which have direct BH mass measurements from dynamical modelling. Additionally, they were able to determine a best-fitting relation that estimates a BH mass from its X-ray and radio luminosities. If we use this relation, we find that the masses predicted by GCM2009 are on average larger than those predicted by the $M_{\rm BH}-\sigma$ and $M_{\rm BH}-M_{K}$ relations, although the scatter remains large. However, if we use the best-fitting relation they find when excluding Seyfert galaxies, we find that the masses predicted by GCM2009 are systematically larger than those predicted by the $M_{\rm BH}-\sigma$ and $M_{\rm BH}-M_{K}$ relations, for all 25 of our BCGs except for the BCG in A2261. Seyfert galaxies are thought to have higher-accretion rates, and are therefore not in a state equivalent to the low-accreting LH state of X-ray binaries.

\section{The existence of UMBHs in BCGs}

If BCGs truly follow the FP of BH activity and that our X-ray and radio core luminosities have not been underestimated or overestimated respectively, then our results imply that the intrinsic masses of the BHs are higher than those predicted from the $M_{\rm BH}-\sigma$ and $M_{\rm BH}-M_{K}$ correlations. 

The sample of MH2007 contains 9 BCGs, 2 of which have precise BH mass measurements from dynamical modelling. The first is Cygnus A, which has a measured BH mass of $2.5\times10^{9}{\rm M_\odot}$ and a predicted BH mass from the $K$ band relation of $2\times10^{9}{\rm M_\odot}$, both of which are consistent with one another. However, the measured BH mass for M87 is significantly higher \cite[$3\times10^{9}{\rm M_\odot}$, MH2007; $6.6\times10^{9}{\rm M_\odot}$,][]{Geb2011729} than that predicted from the $K$ band relation ($0.8\times10^{9}{\rm M_\odot}$). The offset is such that it agrees with our calculations.  

For the remaining 7 BCGs in MH2007, our results imply that the true BH masses lie between $10^9{\rm M_\odot}$ and $10^{10}{\rm M_\odot}$. However, for the 16 BCGs in Table \ref{tab1} that have no detectable X-ray nucleus and that lie in some of the most extreme clusters of galaxies (i.e. some of the most massive and strong cool core), if we apply the average mass increase of ${\log}{\Delta}M_{\rm{BH}}=0.8$, then our results imply that the true BH masses lie between $8\times10^{9}{\rm M_\odot}$ and $2\times10^{10}{\rm M_\odot}$. Furthermore, if we apply the average mass increase derived when including the BCGs from MH2007 (${\log}{\Delta}M_{\rm{BH}}=1.0$, see Table 2), then the majority of these BHs are ultramassive with $M_{\rm BH}=(1-4)\times10^{10}{\rm M_\odot}$.

The existence of UMBHs in BCGs has recently been confirmed by McConnell et al. (2011a)\nocite{McC2011480} who find a $2.1\times10^{10}{\rm M_{\odot}}$ BH in NGC 4889. Yet, NGC 4889 lies in a fairly average cluster with no cool core associated with it. It therefore does not require an active nucleus to prevent the surrounding gas from cooling. On the other hand, the central BHs in massive and strong cool core clusters with $L_{\rm X}>10^{45}$ erg/s and $t_{\rm cool}<3$ Gyr (such as the majority in Table \ref{tab1}) must be injecting extreme mechanical energies into their surrounding medium to prevent the gas from cooling, on the order of $10^{44-45}\ergps$. Out of all BCGs, these are where the BH must have accreted a substantial amount of mass to power the outflows ($>10^{9}M_{\odot}$). Since these BCGs also lie in the most massive clusters, and BCG mass scales with cluster mass, they should host the most massive BHs compared to other BCGs. The most massive BHs should therefore reside in these massive and strong cool core clusters, and our results support this claim, while predicting that many will have UMBHs at their centres. If confirmed, our results will have important ramifications for the formation and evolution of BHs across cosmic time \citep[][]{Lau2007662,McN2009698,Nat2009393,Hla2011,McN2011727,Hla2012421}.

It is also important to note that BCGs have undergone the most dramatic environments, subject to major mergers in the past and extreme AGN outflows for the past several Gigayears. It is not surprising then, that they would have had the opportunity to grow to such masses. There are two possible scenarios in which BHs can grow to be ultramassive. The first is through hierarchical mergers, as supported by numerical calculations \citep[e.g.][ who predict the existence of a rare population of UMBHs in the local Universe]{Yoo2007667}. The second is from high redshift ``seeds" and is based on the observation that quasars exist from as early as $z$ of about $6$ \citep[][]{Vik2011474,Fan2006132}. UMBHs can therefore form from these high redshift ``seeds", and evolve into present-day BCGs, which are the most massive galaxies of the local Universe.  \citet{Nat2009393} argue however that although UMBHs may exist, the maximum mass they can reach is around $\sim10^{10}{\rm M_\odot}$.

\section{Concluding remarks}
We have identified two possibilities as to why BCGs seem to be systematically offset from the FP of BH activity, assuming that our X-ray and radio core luminosities are correct. The first is that the BHs in BCGs are supermassive ($M_{\rm BH}\sim10^9{\rm M_\odot}$), but operate differently from other BHs and lie in a particular place in the FP of BH activity. The second is that these BHs follow the FP of BH activity but not the standard $M_{\rm BH}-\sigma$ and $M_{\rm BH}-M_{K}$ correlations, thus predicting that many of the BHs in massive and strong cool core clusters are ultramassive ($M_{\rm BH}>10^{10}{\rm M_\odot}$). Our results therefore carry significant implications for the formation and evolution of BHs, as well as the connection between the central BH and its host galaxy. Only by obtaining direct BH masses from dynamical modelling of the kinematics, as opposed to relying on known correlations, can we determine if these extreme BHs are truly ultramassive.

\section*{Acknowledgments}
JHL recognizes the support given by the Cambridge Trusts, Natural Sciences and Engineering Research Council of Canada (NSERC), as well as the Fonds Quebecois de la Recherche sur la Nature et les Technologies (FQRNT). ACF thanks the Royal Society. 

\label{lastpage}
\bibliographystyle{mn2e}
\bibliography{bibli}

\begin{thebibliography}{}

\bibitem[\protect\citeauthoryear{{Bell}, {McIntosh}, {Katz} \&
  {Weinberg}}{{Bell} et~al.}{2003}]{Bel2003149}
{Bell} E.~F.,  {McIntosh} D.~H.,  {Katz} N.,    {Weinberg} M.~D.,  2003, \apjs,
  149, 289

\bibitem[\protect\citeauthoryear{{B{\^i}rzan}, {Rafferty}, {McNamara}, {Wise}
  \& {Nulsen}}{{B{\^i}rzan} et~al.}{2004}]{Bir2004607}
{B{\^i}rzan} L.,  {Rafferty} D.~A.,  {McNamara} B.~R.,  {Wise} M.~W.,
  {Nulsen} P.~E.~J.,  2004, \apj, 607, 800

\bibitem[\protect\citeauthoryear{{Blanton} \& {Roweis}}{{Blanton} \&
  {Roweis}}{2007}]{Bla2007133}
{Blanton} M.~R.,  {Roweis} S.,  2007, \aj, 133, 734

\bibitem[\protect\citeauthoryear{{Carilli}, {Perley}, {Dreher} \&
  {Leahy}}{{Carilli} et~al.}{1991}]{Car1991383}
{Carilli} C.~L.,  {Perley} R.~A.,  {Dreher} J.~W.,    {Leahy} J.~P.,  1991,
  \apj, 383, 554

\bibitem[\protect\citeauthoryear{{Churazov}, {Sazonov}, {Sunyaev}, {Forman},
  {Jones} \& {B{\"o}hringer}}{{Churazov} et~al.}{2005}]{Chu2005363}
{Churazov} E.,  {Sazonov} S.,  {Sunyaev} R.,  {Forman} W.,  {Jones} C.,
  {B{\"o}hringer} H.,  2005, \mnras, 363, L91

\bibitem[\protect\citeauthoryear{{Coble}, {Bonamente}, {Carlstrom}, {Dawson},
  {Hasler}, {Holzapfel}, {Joy}, {La Roque}, {Marrone} \& {Reese}}{{Coble}
  et~al.}{2007}]{Cob2008134}
{Coble} K.,  {Bonamente} M.,  {Carlstrom} J.~E.,  {Dawson} K.,  {Hasler} N.,
  {Holzapfel} W.,  {Joy} M.,  {La Roque} S.,  {Marrone} D.~P.,    {Reese}
  E.~D.,  2007, \aj, 134, 897

\bibitem[\protect\citeauthoryear{{Dalla Bont{\`a}}, {Ferrarese}, {Corsini},
  {Miralda-Escud{\'e}}, {Coccato}, {Sarzi}, {Pizzella} \& {Beifiori}}{{Dalla
  Bont{\`a}} et~al.}{2009}]{Dal2009690}
{Dalla Bont{\`a}} E.,  {Ferrarese} L.,  {Corsini} E.~M.,  {Miralda-Escud{\'e}}
  J.,  {Coccato} L.,  {Sarzi} M.,  {Pizzella} A.,    {Beifiori} A.,  2009,
  \apj, 690, 537

\bibitem[\protect\citeauthoryear{{Dunn} \& {Fabian}}{{Dunn} \&
  {Fabian}}{2006}]{Dun2006373}
{Dunn} R.~J.~H.,  {Fabian} A.~C.,  2006, \mnras, 373, 959

\bibitem[\protect\citeauthoryear{{Dunn} \& {Fabian}}{{Dunn} \&
  {Fabian}}{2008}]{Dun2008385}
{Dunn} R.~J.~H.,  {Fabian} A.~C.,  2008, \mnras, 385, 757

\bibitem[\protect\citeauthoryear{{Fabian}, {Sanders}, {Allen}, {Crawford},
  {Iwasawa}, {Johnstone}, {Schmidt} \& {Taylor}}{{Fabian}
  et~al.}{2003}]{Fab2003344}
{Fabian} A.~C.,  {Sanders} J.~S.,  {Allen} S.~W.,  {Crawford} C.~S.,  {Iwasawa}
  K.,  {Johnstone} R.~M.,  {Schmidt} R.~W.,    {Taylor} G.~B.,  2003, \mnras,
  344, L43

\bibitem[\protect\citeauthoryear{{Fabian}, {Sanders}, {Taylor}, {Allen},
  {Crawford}, {Johnstone} \& {Iwasawa}}{{Fabian} et~al.}{2006}]{Fab2006366}
{Fabian} A.~C.,  {Sanders} J.~S.,  {Taylor} G.~B.,  {Allen} S.~W.,  {Crawford}
  C.~S.,  {Johnstone} R.~M.,    {Iwasawa} K.,  2006, \mnras, 366, 417

\bibitem[\protect\citeauthoryear{{Falcke}, {K{\"o}rding} \& {Markoff}}{{Falcke}
  et~al.}{2004}]{Fal2004414}
{Falcke} H.,  {K{\"o}rding} E.,    {Markoff} S.,  2004, \aap, 414, 895

\bibitem[\protect\citeauthoryear{{Fan}, {Strauss}, {Becker}, {White}, {Gunn},
  {Knapp}, {Richards}, {Schneider}, {Brinkmann} \& {Fukugita}}{{Fan}
  et~al.}{2006}]{Fan2006132}
{Fan} X.,  {Strauss} M.~A.,  {Becker} R.~H.,  {White} R.~L.,  {Gunn} J.~E.,
  {Knapp} G.~R.,  {Richards} G.~T.,  {Schneider} D.~P.,  {Brinkmann} J.,
  {Fukugita} M.,  2006, \aj, 132, 117

\bibitem[\protect\citeauthoryear{{Fender}}{{Fender}}{2001}]{Fen2001322}
{Fender} R.~P.,  2001, \mnras, 322, 31

\bibitem[\protect\citeauthoryear{{Forman}, {Nulsen}, {Heinz}, {Owen}, {Eilek},
  {Vikhlinin}, {Markevitch}, {Kraft}, {Churazov} \& {Jones}}{{Forman}
  et~al.}{2005}]{For2005635}
{Forman} W.,  {Nulsen} P.,  {Heinz} S.,  {Owen} F.,  {Eilek} J.,  {Vikhlinin}
  A.,  {Markevitch} M.,  {Kraft} R.,  {Churazov} E.,    {Jones} C.,  2005,
  \apj, 635, 894

\bibitem[\protect\citeauthoryear{{Gebhardt}, {Adams}, {Richstone}, {Lauer},
  {Faber}, {G{\"u}ltekin}, {Murphy} \& {Tremaine}}{{Gebhardt}
  et~al.}{2011}]{Geb2011729}
{Gebhardt} K.,  {Adams} J.,  {Richstone} D.,  {Lauer} T.~R.,  {Faber} S.~M.,
  {G{\"u}ltekin} K.,  {Murphy} J.,    {Tremaine} S.,  2011, \apj, 729, 119

\bibitem[\protect\citeauthoryear{{Giacintucci}, {Markevitch}, {Brunetti},
  {Cassano} \& {Venturi}}{{Giacintucci} et~al.}{2011}]{Gia2011525}
{Giacintucci} S.,  {Markevitch} M.,  {Brunetti} G.,  {Cassano} R.,    {Venturi}
  T.,  2011, \aap, 525, L10+

\bibitem[\protect\citeauthoryear{{Gioia}, {Maccacaro}, {Schild}, {Wolter},
  {Stocke}, {Morris} \& {Henry}}{{Gioia} et~al.}{1990}]{Gio199072}
{Gioia} I.~M.,  {Maccacaro} T.,  {Schild} R.~E.,  {Wolter} A.,  {Stocke} J.~T.,
   {Morris} S.~L.,    {Henry} J.~P.,  1990, \apjs, 72, 567

\bibitem[\protect\citeauthoryear{{Graham}}{{Graham}}{2007}]{Gra2007379}
{Graham} A.~W.,  2007, \mnras, 379, 711

\bibitem[\protect\citeauthoryear{{G{\"u}ltekin}, {Cackett}, {Miller}, {Di
  Matteo}, {Markoff} \& {Richstone}}{{G{\"u}ltekin} et~al.}{2009}]{Gul2009706}
{G{\"u}ltekin} K.,  {Cackett} E.~M.,  {Miller} J.~M.,  {Di Matteo} T.,
  {Markoff} S.,    {Richstone} D.~O.,  2009, \apj, 706, 404

\bibitem[\protect\citeauthoryear{{Hlavacek-Larrondo} \&
  {Fabian}}{{Hlavacek-Larrondo} \& {Fabian}}{2011}]{Hla2011}
{Hlavacek-Larrondo} J.,  {Fabian} A.~C.,  2011, \mnras, 413, 313

\bibitem[\protect\citeauthoryear{{Hlavacek-Larrondo}, {Fabian}, {Edge},
  {Ebeling}, {Sanders}, {Hogan} \& {Taylor}}{{Hlavacek-Larrondo}
  et~al.}{2012}]{Hla2012421}
{Hlavacek-Larrondo} J.,  {Fabian} A.~C.,  {Edge} A.~C.,  {Ebeling} H.,
  {Sanders} J.~S.,  {Hogan} M.~T.,    {Taylor} G.~B.,  2012, \mnras, 421, 1360

\bibitem[\protect\citeauthoryear{{Houghton}, {Magorrian}, {Sarzi}, {Thatte},
  {Davies} \& {Krajnovi{\'c}}}{{Houghton} et~al.}{2006}]{Hou2006367}
{Houghton} R.~C.~W.,  {Magorrian} J.,  {Sarzi} M.,  {Thatte} N.,  {Davies}
  R.~L.,    {Krajnovi{\'c}} D.,  2006, \mnras, 367, 2

\bibitem[\protect\citeauthoryear{{Jarrett}, {Chester}, {Cutri}, {Schneider},
  {Skrutskie} \& {Huchra}}{{Jarrett} et~al.}{2000}]{Jar2000119}
{Jarrett} T.~H.,  {Chester} T.,  {Cutri} R.,  {Schneider} S.,  {Skrutskie} M.,
    {Huchra} J.~P.,  2000, \aj, 119, 2498

\bibitem[\protect\citeauthoryear{{K{\"o}rding}, {Falcke} \&
  {Corbel}}{{K{\"o}rding} et~al.}{2006}]{Kor2006456}
{K{\"o}rding} E.,  {Falcke} H.,    {Corbel} S.,  2006, \aap, 456, 439

\bibitem[\protect\citeauthoryear{{Lauer}, {Faber}, {Richstone}, {Gebhardt},
  {Tremaine}, {Postman}, {Dressler}, {Aller}, {Filippenko}, {Green}, {Ho},
  {Kormendy}, {Magorrian} \& {Pinkney}}{{Lauer} et~al.}{2007}]{Lau2007662}
{Lauer} T.~R.,  {Faber} S.~M.,  {Richstone} D.,  {Gebhardt} K.,  {Tremaine} S.,
   {Postman} M.,  {Dressler} A.,  {Aller} M.~C.,  {Filippenko} A.~V.,  {Green}
  R.,  {Ho} L.~C.,  {Kormendy} J.,  {Magorrian} J.,    {Pinkney} J.,  2007,
  \apj, 662, 808

\bibitem[\protect\citeauthoryear{{Macchetto}, {Marconi}, {Axon}, {Capetti},
  {Sparks} \& {Crane}}{{Macchetto} et~al.}{1997}]{Mac1997489}
{Macchetto} F.,  {Marconi} A.,  {Axon} D.~J.,  {Capetti} A.,  {Sparks} W.,
  {Crane} P.,  1997, \apj, 489, 579

\bibitem[\protect\citeauthoryear{{McConnell}, {Ma}, {Gebhardt}, {Wright},
  {Murphy}, {Lauer}, {Graham} \& {Richstone}}{{McConnell}
  et~al.}{011a}]{McC2011480}
{McConnell} N.~J.,  {Ma} C.-P.,  {Gebhardt} K.,  {Wright} S.~A.,  {Murphy}
  J.~D.,  {Lauer} T.~R.,  {Graham} J.~R.,    {Richstone} D.~O.,  2011a, \nat,
  480, 215

\bibitem[\protect\citeauthoryear{{McConnell}, {Ma}, {Graham}, {Gebhardt},
  {Lauer}, {Wright} \& {Richstone}}{{McConnell} et~al.}{011b}]{McC2011728}
{McConnell} N.~J.,  {Ma} C.-P.,  {Graham} J.~R.,  {Gebhardt} K.,  {Lauer}
  T.~R.,  {Wright} S.~A.,    {Richstone} D.~O.,  2011b, \apj, 728, 100

\bibitem[\protect\citeauthoryear{{McNamara}, {Kazemzadeh}, {Rafferty},
  {B{\^i}rzan}, {Nulsen}, {Kirkpatrick} \& {Wise}}{{McNamara}
  et~al.}{2009}]{McN2009698}
{McNamara} B.~R.,  {Kazemzadeh} F.,  {Rafferty} D.~A.,  {B{\^i}rzan} L.,
  {Nulsen} P.~E.~J.,  {Kirkpatrick} C.~C.,    {Wise} M.~W.,  2009, \apj, 698,
  594

\bibitem[\protect\citeauthoryear{{McNamara} \& {Nulsen}}{{McNamara} \&
  {Nulsen}}{2007}]{Mcn200745}
{McNamara} B.~R.,  {Nulsen} P.~E.~J.,  2007, \araa, 45, 117

\bibitem[\protect\citeauthoryear{{McNamara}, {Rohanizadegan} \&
  {Nulsen}}{{McNamara} et~al.}{2011}]{McN2011727}
{McNamara} B.~R.,  {Rohanizadegan} M.,    {Nulsen} P.~E.~J.,  2011, \apj, 727,
  39

\bibitem[\protect\citeauthoryear{{Merloni} \& {Heinz}}{{Merloni} \&
  {Heinz}}{2007}]{Mer2007381}
{Merloni} A.,  {Heinz} S.,  2007, \mnras, 381, 589

\bibitem[\protect\citeauthoryear{{Merloni}, {Heinz} \& {di Matteo}}{{Merloni}
  et~al.}{2003}]{Mer2003345}
{Merloni} A.,  {Heinz} S.,    {di Matteo} T.,  2003, \mnras, 345, 1057

\bibitem[\protect\citeauthoryear{{Mukai}}{{Mukai}}{1993}]{Mukai1993}
{Mukai} K., , 1993, {Legacy}, 3, 21

\bibitem[\protect\citeauthoryear{{Natarajan} \& {Treister}}{{Natarajan} \&
  {Treister}}{2009}]{Nat2009393}
{Natarajan} P.,  {Treister} E.,  2009, \mnras, 393, 838

\bibitem[\protect\citeauthoryear{{Plotkin}, {Markoff}, {Kelly}, {K{\"o}rding}
  \& {Anderson}}{{Plotkin} et~al.}{2012}]{Plo2012419}
{Plotkin} R.~M.,  {Markoff} S.,  {Kelly} B.~C.,  {K{\"o}rding} E.,
  {Anderson} S.~F.,  2012, \mnras, 419, 267

\bibitem[\protect\citeauthoryear{{Rafferty}, {McNamara}, {Nulsen} \&
  {Wise}}{{Rafferty} et~al.}{2006}]{Raf2006652}
{Rafferty} D.~A.,  {McNamara} B.~R.,  {Nulsen} P.~E.~J.,    {Wise} M.~W.,
  2006, \apj, 652, 216

\bibitem[\protect\citeauthoryear{{Sanders} \& {Fabian}}{{Sanders} \&
  {Fabian}}{2007}]{San2007381}
{Sanders} J.~S.,  {Fabian} A.~C.,  2007, \mnras, 381, 1381

\bibitem[\protect\citeauthoryear{{Schlegel}, {Finkbeiner} \&
  {Davis}}{{Schlegel} et~al.}{1998}]{Sch1998500}
{Schlegel} D.~J.,  {Finkbeiner} D.~P.,    {Davis} M.,  1998, \apj, 500, 525

\bibitem[\protect\citeauthoryear{{Stocke}, {Morris}, {Gioia}, {Maccacaro},
  {Schild}, {Wolter}, {Fleming} \& {Henry}}{{Stocke} et~al.}{1991}]{Sto199176}
{Stocke} J.~T.,  {Morris} S.~L.,  {Gioia} I.~M.,  {Maccacaro} T.,  {Schild} R.,
   {Wolter} A.,  {Fleming} T.~A.,    {Henry} J.~P.,  1991, \apjs, 76, 813

\bibitem[\protect\citeauthoryear{{Stott}, {Edge}, {Smith}, {Swinbank} \&
  {Ebeling}}{{Stott} et~al.}{2008}]{Sto2008384}
{Stott} J.~P.,  {Edge} A.~C.,  {Smith} G.~P.,  {Swinbank} A.~M.,    {Ebeling}
  H.,  2008, \mnras, 384, 1502

\bibitem[\protect\citeauthoryear{{Vikhlinin}}{{Vikhlinin}}{2011}]{Vik2011474}
{Vikhlinin} A.,  2011, \nat, 474, 293

\bibitem[\protect\citeauthoryear{{von der Linden}, {Best}, {Kauffmann} \&
  {White}}{{von der Linden} et~al.}{2007}]{von2007379}
{von der Linden} A.,  {Best} P.~N.,  {Kauffmann} G.,    {White} S.~D.~M.,
  2007, \mnras, 379, 867

\bibitem[\protect\citeauthoryear{{Yoo}, {Miralda-Escud{\'e}}, {Weinberg},
  {Zheng} \& {Morgan}}{{Yoo} et~al.}{2007}]{Yoo2007667}
{Yoo} J.,  {Miralda-Escud{\'e}} J.,  {Weinberg} D.~H.,  {Zheng} Z.,    {Morgan}
  C.~W.,  2007, \apj, 667, 813

\end{thebibliography}

\end{document}